\documentclass[12pt,a4paper]{article}
\usepackage{amsmath,amssymb,psfrag,graphicx}
\usepackage[nosort]{cite}

\begin{document}
\input{epsf}
\topmargin 0pt
\oddsidemargin 5mm
\headheight 0pt
\headsep 0pt
\topskip 9mm
\pagestyle{empty}

\newcommand{\beq}{\begin{equation}}
\newcommand{\eeq}{\end{equation}}
\newcommand{\bea}{\begin{eqnarray}}
\newcommand{\eea}{\end{eqnarray}}
\newcommand{\rf}[1]{(\ref{#1})}
\newcommand{\C}[1]{{\cal #1}}
\newcommand{\pa}{\partial}
\newcommand{\nn}{\nonumber}

\newcommand{\e}{\mbox{e}}
\renewcommand{\d}{\mbox{d}}
\newcommand{\g}{\gamma}
\newcommand{\G}{\Gamma}
\renewcommand{\l}{\lambda}
\renewcommand{\L}{\Lambda}
\renewcommand{\b}{\beta}
\renewcommand{\a}{\alpha}
\newcommand{\n}{\nu}
\newcommand{\m}{\mu}
\newcommand{\Tr}{\mbox{Tr}}
\newcommand{\E}{\mbox{E(q)}}
\newcommand{\Ee}{\mbox{E}}
\newcommand{\K}{\mbox{K(q)}}
\newcommand{\Kk}{\mbox{K}}
\newcommand{\ep}{\varepsilon}
\newcommand{\om}{\omega}
\newcommand{\del}{\delta}
\newcommand{\Del}{\Delta}
\newcommand{\sg}{\sigma}
\newcommand{\vph}{\varphi}
\newcommand{\sn}{\mbox{sn}}
\newcommand{\dn}{\mbox{dn}}
\newcommand{\cn}{\mbox{cn}}

\newcommand{\oh}{\frac{1}{2}}
\newcommand{\oq}{\frac{1}{4}}
\newcommand{\dg}{\dagger}

\newcommand{\tr}{\mbox{Tr}\;}
\newcommand{\ra}{\right\rangle}
\newcommand{\la}{\left\langle}
\newcommand{\prt}{\partial}
\newcommand{\mi}{\!-\!}
\newcommand{\equ}{\!=\!}
\newcommand{\pl}{\!+\!}

\newcommand{\cD}{{\cal D}}
\newcommand{\cS}{{\cal S}}
\newcommand{\cM}{{\cal M}}
\newcommand{\cK}{{\cal K}}
\newcommand{\cT}{{\cal T}}
\newcommand{\cN}{{\cal N}}
\newcommand{\cL}{{\cal L}}
\newcommand{\cO}{{\cal O}}
\newcommand{\cR}{{\cal R}}

\newcommand{\tF}{{\tilde{F}}}
\newcommand{\tL}{{\tilde{\L}}}
\newcommand{\tX}{{\tilde{X}}}
\newcommand{\tY}{{\tilde{Y}}}
\newcommand{\tZ}{{\tilde{Z}}}
\newcommand{\ty}{{\tilde{y}}}
\newcommand{\tz}{{\tilde{z}}}
\newcommand{\tg}{{\tilde{g}}}
\newcommand{\tG}{{\tilde{G}}}
\newcommand{\tH}{{\tilde{H}}}
\newcommand{\tT}{{\tilde{T}}}

\newcommand{\SL}{{\sqrt{\L}}}
\newcommand{\tSL}{\sqrt{\tL}}
\newcommand{\FL}{\L^{1/4}}
\newcommand{\bZ}{{\bar{Z}}}
\newcommand{\bX}{{\bar{X}}}

\newcommand{\remark}[1]{{\renewcommand{\bfdefault}{b}\textbf{\mathversion{bold}#1}}}

\begin{flushright}
{\sc\footnotesize NORDITA-2007-14}\\
\end{flushright}
\vspace*{100pt}

\begin{center}

{\large \bf { The strong coupling limit of the scaling function
from the quantum string Bethe Ansatz}}

\vspace*{26pt}

{\sl P.Y.\ Casteill}$\,^{a}$ and
{\sl C.\ Kristjansen}$\,^{b}$

\vspace{10pt}
\vspace{10pt}

$^a$~The Niels Bohr Institute, Copenhagen University\\
Blegdamsvej 17, DK-2100 Copenhagen \O , Denmark.\\
\vspace{10pt}

$^b$~The Niels Bohr Institute and NORDITA, Copenhagen University\\
Blegdamsvej 17, DK-2100 Copenhagen \O , Denmark.\\
\vspace{10pt}

\vspace{10pt}

\vspace{20pt}

\end{center}

\begin{abstract}
\noindent

Using the quantum string Bethe ansatz we derive the one-loop energy
of a folded string rotating with angular momenta $(S,J)$ in
$AdS_3\times S^1\subset AdS_5\times S^5$ in the limit $1\ll J\ll S$,
$z=\sqrt{\lambda} \log(S/J) /(\pi J)$ fixed. The one-loop energy is a sum of
two contributions, one originating from the Hernandez-Lopez
phase and another one being due to spin chain finite size effects. We find
a result which at the functional level exactly matches the result of
a string theory computation. Expanding the result for large $z$ we obtain
the strong coupling limit of the scaling function for low twist, high
spin operators of the $SL(2)$ sector of ${\cal N}=4$ SYM. In particular
we recover the famous $-\frac{3\log(2)}{\pi}$. Its appearance is a
result of non-trivial cancellations between
the finite size effects and the Hernandez-Lopez correction.

\vfill

\vspace*{0.4cm}

\vspace*{0.2cm}

\vspace*{0.2cm}
\noindent
arXiv:0705.0890 [hep-th]

\end{abstract}

\newpage

\pagestyle{plain}

\setcounter{page}{1}

\newcommand{\ft}[2]{{\textstyle\frac{#1}{#2}}}
\newcommand{\ii}{\mathrm{i}}
\newcommand{\dd}{{\mathrm{d}}}
\newcommand{\nnb}{\nonumber}

\section{Introduction}

 Due to recent years discovery of integrable models underlying the
spectral problems of both ${\cal N}=4$ SYM~\cite{Minahan:2002ve}
and type IIB string theory on $AdS_5\times S^5$~\cite{Mandal:2002fs}
the spectral part of the AdS/CFT conjecture~\cite{Maldacena:1997re} can now be
stated in a very pointed manner. Namely, the conjecture simply says that
the $S$-matrix of the respective integrable models must
agree~\cite{Staudacher:2004tk}.
Furthermore, the common symmetry group of the two theories constrains
the S-matrix up to a phase factor~\cite{Beisert:2005tm}.
The formulation of the conjecture can thus
be further sharpened to the statement that the phase factors of respectively
${\cal N}=4$ SYM and type IIB string theory on $AdS_5\times S^5$ should be
identical.

Based on educated guessing, phase factors for both ${\cal N}=4$ SYM and
type IIB string theory on $AdS_5\times S^5$ have been put forward.
In accordance with the strong-weak coupling nature of the AdS/CFT
correspondence the gauge theory phase factor~\cite{Beisert:2006ez} is
given as an infinite series in the 't Hooft coupling constant $\lambda$
whereas the string theory phase factor~\cite{Beisert:2006ib}
is given as an asymptotic expansion in
$\frac{1}{\sqrt{\lambda}}$.
There exist arguments that the string theory asymptotic
expansion for large $\lambda$ can originate from the same function as defined
by the gauge theory perturbative expansion which has a finite radius of
convergence~\cite{Beisert:2006ez}. However, both phase factors are rather
involved functions and it would be reassuring to see an example of a
simple observable which can be extrapolated smoothly from weak to
strong coupling.

A candidate for such an observable is the universal scaling function or
cusp anomalous dimension, $f(g)$ where $g^2=\frac{\lambda}{8\pi^2}$. 
It is related to the anomalous dimension
of low twist operators of ${\cal N}=4$ SYM of the type
\beq
\label{twist}{\cal O}=\Tr (D^S Z^J+\ldots).
\eeq
Here $D$ is a light cone derivative, $Z$ is a complex scalar, $S$ is the
space-time spin and $J$ is denoted as the twist. For leading twist, i.e.\
$J=2$, it is well-known that the anomalous dimension $\Delta$
of such an operator for large values of the
spin grows logarithmically with the spin
\beq
\Delta- S=f(g)\log(S),\label{scaling}
\hspace{0.7cm} S\rightarrow \infty,
\eeq
where $f(g)$ can be expanded perturbatively in $g$.
The scaling function has the appealing feature that, as opposed to other
observables one could think of, it depends only on one parameter $g$.
For instance, it is not polluted by any additional $J$-dependence.
The function $f(g)$ has been determined by solid field theory calculations
up to and including four-loop
order~\cite{Bern:2006ew}.
Furthermore, starting from
the asymptotic
gauge theory Bethe equations~\cite{Beisert:2005fw},
inserting the conjectured gauge theory phase
factor~\cite{Beisert:2006ez}
and taking a large-$S$ limit it has been possible to derive an equation
which determines $f(g)$ to all
orders in $g$~\cite{Eden:2006rx}.
This equation, known as the BES equation, correctly reproduces
the known first four orders in $g^2$.
Its derivation, however, relies
on the assumption that the scaling function is the same for all operators with
a {\em finite} value of the twist and that at the same time it is
permitted to take $J$ sufficiently large so that the asymptotic Bethe
equations are correct.

On the string theory side a low twist, high spin operator corresponds to a
folded string rotating with angular momentum $S$ on
$AdS_3\subset AdS_5\times S^5$~\cite{Gubser:2002tv}.
The energy of such a string has an expansion
for large $\lambda$ which reads
\beq
E=\left( \frac{\sqrt{\lambda}}{\pi}-\frac{3\log(2)}{\pi}+
{\cal O}\left(\frac{1}{\sqrt{\lambda}}\right)\right) \log S, \hspace{0.7cm}
S\rightarrow\infty.
\label{AdS}
\eeq
Here the first term follows from semi-classical analysis~\cite{Gubser:2002tv}
and the second one from a one-loop computation~\cite{Frolov:2002av}.
Deriving this result from the Bethe equations
would yield a very comforting confirmation of both
the integrability approach as well as of the AdS/CFT conjecture itself.
However, the strong coupling analysis of the BES equation has proved
hard. For the moment only the leading semi-classical contribution has
been derived from the BES equation by analytic
means~\cite{Kotikov:2006ts,Alday:2007qf,Kostov:2007kx}.
By numerical analysis of the equation both
the leading~\cite{Benna:2006nd,Beccaria:2007tk} and the next to leading
order term~\cite{Benna:2006nd} can be reproduced with high accuracy.
Furthermore, it is possible
to predict numerically the next term in the expansion which would result
from a string theory two-loop computation~\cite{Benna:2006nd}.
 In the present paper we shall consider an alternative way of obtaining
an expansion \`{a} la~\rf{AdS} by Bethe equation techniques.

An operator of the type~\rf{twist} for which $J$ is not finite
has a  string
theory dual which
in addition to the angular momentum, $S$ on $AdS_3$ carries an angular
momentum $J$ on $S^1\subset S^5$. For such a string, considering
the situation
\beq
1\ll J \ll S, \hspace{0.7cm} z\equiv \frac{\sqrt{\lambda}}{\pi J}
\log\left(\frac{S}{J}\right),\hspace{0.3cm} \mbox{fixed},
\label{limit}
\eeq
one finds that the semi-classical~\cite{Belitsky:2006en} as well as the one loop
energy~\cite{Frolov:2006qe}
can be written down in a closed form as a function of $z$. Furthermore,
the formula obtained for the string energy
interpolates smoothly between small and large values
of $z$ and the large-$z$ expansion looks as~\rf{AdS} just
with the replacement $\log S\rightarrow \log(\frac{S}{J})$. We shall
discuss this string solution and the various expansions of its energy
in more detail shortly. Subsequently, we will show  how to
reproduce the precise
functional dependence of the string energy on $z$ from the string
Bethe equations.
In particular, we will derive by analytic means the celebrated
$-\frac{3\log(2)}{\pi}$. Our starting point will be the
asymptotic Bethe equations, whose application is now more justified
since we take $J\gg 1$, supplemented with the conjectured string phase factor.
The classical string energy
as a function of $z$ is obtained almost immediately by considering
only the AFS phase~\cite{Arutyunov:2004vx} whereas the one-loop
energy requires more work. For one we have to take into account the
HL-correction~\cite{Hernandez:2006tk} to the phase and secondly we have to
consider spin-chain finite size effects~\cite{Beisert:2005mq}.
 As we shall see we are able to
determine the contribution from each of these effects exactly as
a function of $z$. The $-\frac{3\log(2)}{\pi}$ results from a non-trivial
cancellation between the two types of terms as $z\rightarrow \infty$.

We start in section~\ref{foldedstring} by  recalling from
reference~\cite{Frolov:2006qe}
the description of the folded string rotating
on $AdS_3\times S^1\subset AdS_5\times S^5$ in the limit given
by eqn.~\rf{limit}. In section~\ref{Betheequations} we write down the relevant
string Bethe equations and perform the necessary expansions. After that,
in sections~\ref{semiclassical} and~\ref{oneloop}, we extract from these
respectively the semi-classical and the one-loop energy.
Finally, section~\ref{Conclusion} contains our conclusion.

\section{The folded string spinning on $AdS_3\times S^1$ \label{foldedstring}}

A folded string living in $AdS_5\times S^5$ and carrying large
angular momenta $S$ and $J$ on respectively $AdS_3$ and $S^1$ is a system
which has successfully been studied in the semi-classical approximation.
Hence, its
classical energy was determined in~\cite{Frolov:2002av}. The expression
for the energy simplifies considerably in the limit given
in eqn.~\rf{limit}, i.e.~\cite{Belitsky:2006en}
\beq
1\ll J\ll S, \hspace{0.7cm} z\equiv \frac{\sqrt{\lambda}}{\pi J}
\log \left( \frac{S}{J}\right) \hspace{0.3cm} \mbox{fixed}.
\label{limit2}
\eeq
One finds
\beq
E_0 = S+J \sqrt{1+z^2}. \label{E0}
\eeq
Expanding for large $z$ we get
\beq
E_0(z\gg 1)= S+\frac{\sqrt{\lambda}}{\pi} \log \left(\frac{S}{J}\right)+\ldots.
\eeq
Here we notice the leading strong coupling term announced
earlier, cf.\ eqn.~\rf{AdS}.
For $z\ll 1$ one recovers what is known as the fast spinning string
solution~\cite{Frolov:2003qc}
\beq
E_0(z\ll 1)=S+J+\frac{\lambda}{2\pi^2 J}\log^2 \left(\frac{S}{J}\right)
-\frac{\lambda^2}{8\pi^4 J^3} \log^4 \left(\frac{S}{J}\right)
+\frac{\lambda^3}{16\pi^6 J^5} \log^6 \left(\frac{S}{J}\right)+\ldots.
\label{E0weak}
\eeq
The first logarithmic term was reproduced in the Bethe ansatz approach
in~\cite{Beisert:2003ea} and the second one is contained in the work
in~\cite{Belitsky:2006en}. Later, we shall show that when the limit~\rf{limit2}
is imposed from the beginning in the all Loop Bethe ansatz,
the exact square root formula immediately appears. Recently, an expression
for the one-loop contribution to the energy in the same limit was
derived~\cite{Frolov:2006qe}. The result reads
\bea
\lefteqn{E_1=\frac{J}{\sqrt{\lambda}}
\frac{1}{\sqrt{1+z^2}} \left\{ z\sqrt{1+z^2}-(1+2z^2)
\log\left[z+\sqrt{1+z^2}\right]\right. }\nonumber \\
&&\hspace{1.5cm}\left.-z^2+2(1+z^2)\log(1+z^2)
-(1+2z^2)\log\left[\sqrt{1+2z^2}\right]
 \right\}. \label{E1}
\eea
It is obtained under the further assumption that
\beq
\frac{J}{\sqrt{\lambda}} \sqrt{1+z^2}\gg 1.
\eeq
Expanding~\rf{E1} for small $z$, we get for the fast spinning case
\beq
E_1(z\ll 1)=-\frac{4\lambda}{3\pi^3 J^2} \log^3 \left( \frac{S}{J}\right)
+\frac{4\lambda^2}{5\pi^5 J^4} \log^5 \left( \frac{S}{J}\right)
+\frac{\lambda^{5/2}}{3\pi^6 J^5} \log^6 \left( \frac{S}{J}\right)+\ldots.
\eeq
Taking in stead $z$ to be large, one finds
\beq
E_1(z\gg 1)=- \frac{3\log(2)}{\pi} \log \left(\frac{S}{J}\right)+\ldots.
\eeq
Here we recognize the famous $-\frac{3\log(2)}{\pi}$
coefficient from the large-$\lambda$ expansion~\rf{AdS}.
As we shall explain in the next section, from the Bethe equation perspective
it is natural to separate $E_1$ into a part which is analytic in $\lambda$
for small $\lambda$ and one which is not. Terms which are analytic,
respectively non-analytic, in $\lambda$
originate from terms which are odd, respectively even, in $z$. (The
even terms constitute the first line in eqn.~\rf{E1} and the odd ones
the second line.) Explicitly, we have
\bea
(E_1)_{\hbox{\scriptsize string}}^{\hbox{\scriptsize analytic}}
&=& \frac{J}{\sqrt{\lambda}} \left(
z-\frac{1+2z^2}{\sqrt{1+z^2}}\log \left[ z+\sqrt{1+z^2}\right]\right)
\label{analytic1} \\
&=& \mbox{}-\frac{4\lambda \log^3\left(\frac{S}{J}\right)}{\pi^3 J^2}
\left(\frac{1}{3}-\frac{1}{5}z^2+\frac{16}{105}z^4+\ldots \right),
\label{analytic2} \\
(E_1)^{\hbox{\scriptsize non-analytic}}_{{\hbox{\scriptsize string}}}&=&
\frac{J}{\sqrt{\lambda}} \frac{1}{\sqrt{1+z^2}}
\left(
-z^2+2(1+z^2)\log(1+z^2) \right.  \nonumber\\
&&\left.\mbox{}-(1+2z^2)\log\left[\sqrt{1+2z^2}\right] \right)
\label{nonanalytic1} \\
&=&\frac{\lambda^{5/2} \log^6\left(\frac{S}{J}\right)}{\pi^6 J^5}
\left( \frac{1}{3}-\frac{2}{3}z^2+\frac{43}{40}z^4 +\ldots \right).
\eea
The first term in the expansion~\rf{analytic2} of the analytic
part was
recovered using the one-loop Bethe ansatz in~\cite{Belitsky:2006en}.
Below we shall recover the exact functional expressions~\rf{analytic1}
and~\rf{nonanalytic1}. It is an important point to notice that the appearance
of the $-\frac{3\log(2)}{\pi}$ term for large $z$ is due to non-trivial
cancellations between the analytic and the non-analytic part.
More precisely, we have 
\bea
\left(E_1\right)^{\hbox{\scriptsize analytic}}_{\hbox{\scriptsize string}}& 
\sim & \left(\frac{-2\log(z)+1-2\log(2)}{\pi}\right)
\log\left(\frac{S}{J}\right)
\hspace{0.7cm} \mbox{as} \hspace{0.7cm}
z\rightarrow \infty,\\
\left(E_1\right)^{\hbox{\scriptsize non-analytic}}_{\hbox{\scriptsize string}} &\sim& \left(\frac{2\log(z)-1-\log(2)}{\pi}\right)
\log\left(\frac{S}{J}\right)
\hspace{0.7cm} \mbox{as} \hspace{0.7cm}
z\rightarrow \infty.
\eea

\section{The string Bethe equations\label{Betheequations}}

The spectrum of strings moving on $AdS_3\times S^1\subset AdS_5\times S^5$
is encoded in the Bethe equations of a generalized $\mathfrak{sl}(2)$
spin chain, i.e.

\begin{equation}\label{Betheallloop}
\left(\frac{x_k^+}{x_k^-}\right)^J=
\prod_{j\neq k}^S
\left(\frac{x_k^--x_j^+}{x_k^+-x_j^-}\right)
\frac{1-g^2/2x_k^+x_j^-}{1-g^2/2x_j^+x_k^-}\,
\sigma^2(x_k,x_j),
\end{equation}
Here $S$ and $J$ are representation labels associated with the angular momentum
of the string on respectively $AdS_3$ and $S^1$ and $g$ is the inverse
string tension
\beq
g^2=\frac{\lambda}{8\pi^2}\sim \frac{1}{\alpha'^2}.
\eeq
The indices $j,k$ label elementary excitations and the $x^{\pm}$ variables
are related to the momenta carried by these excitations via
\beq
\exp(\ii\, p)=\frac{x^+}{x^-}.
\eeq
Furthermore, the quantity $\sigma(x_k,x_j)$ is the phase factor, restricted
by symmetry arguments to be of the form~\cite{Beisert:2005wv}
\begin{eqnarray}
\nonumber \sigma(x_k,x_j)&=&e^{\ii\,\theta(x_k,x_j)},\\
\theta(x_k,x_j)&=&\sum_{r=2}^\infty \sum_{s=r+1}^\infty
\left(\frac{g^2}{2}\right)^{(r+s-1)/2}c_{r,s}(g)
\left[q_r(x_k)q_s(x_j)-q_r(x_j)q_s(x_k)\right],\label{AFSphase}
\end{eqnarray}
where the charges $q_r(x)$ (with $r\geq 2$) are defined by
\begin{equation}
\quad q_r(x_k)=\frac{\ii}{r-1}\left(\frac{1}{(x_k^+)^{r-1}}-\frac{1}{(x_k^-)^{r-1}}\right), \hspace{0.7cm}Q_r=\sum_kq_r(x_k).
\end{equation}
In the string theory description, the $c_{r,s}$ coefficients are expected to have an expansion
in $\alpha'\sim \frac{1}{\sqrt{\lambda}}$
\beq
c_{r,s}(\lambda)=c_{r,s}^{(0)}+\frac{1}{\sqrt{\lambda}} c_{r,s}^{(1)}+
\frac{1}{\lambda} c_{r,s}^{(2)}+\ldots, \label{crs}
\eeq
and the string phase factor conjecture~\cite{Beisert:2006ib}
accordingly
involves an explicit conjecture for the $c_{r,s}^{(i)}$. The first two terms
can be determined by comparing to conventional string theory computations
and read~\cite{Arutyunov:2004vx,Hernandez:2006tk}
\bea
c_{r,s}^{(0)}&=&\delta_{s,r+1}, \\
c_{r,s}^{(1)}&=&-4(1-(-1)^{r+s}) \frac{(r-1)(s-1)}{(s+r-2)(s-r)}.\label{coefc1}
\eea
In order to describe proper string states the Bethe equations must be
supplemented by the level matching or momentum condition
\beq
\prod_{k=1}^S \left(\frac{x_k^+}{x_k^-}\right)=1,
\eeq
and finally the string energy is then obtained as
\beq
E=\frac{\lambda}{8\pi^2}Q_2.
\eeq
Now, our aim is to determine the classical and the one-loop energy of
a certain string configuration in the limit given by eqn.~\rf{limit2}.
For that purpose we need to expand the phase factor to two leading
orders in $\alpha'\sim \frac{1}{\sqrt{\lambda}}$., i.e.\ to
take into account $c_{r,s}^{(0)}$ and $c_{r,s}^{(1)}$ above. Correspondingly,
we have to expand all terms to two leading orders in $\frac{1}{J}$.
In order to perform the large-$J$ expansion we need to express the
$x$-variables via a rapidity variable $u$ in the following way
\bea
x^{\pm}&=& x(u\pm  \ii/2), \\
x(u)&=& \frac{u}{2}+\frac{u}{2}\sqrt{1-\frac{2g^2}{u^2}}, \\
u(x)&=& x+\frac{g^2}{2x}.
\eea
We then rescale the variables $x=x(u)$ and $g^2$ in the following way
\bea
x=x(u)&\rightarrow & J\;x, \\
g^2 &\rightarrow & g^2J^2.
\eea
Taking the logarithm of the Bethe equations and expanding
to the relevant order in $J$ and $\lambda$
we obtain
\bea
\lefteqn{-\frac{1}{x_k (1-g^2/(2x_k^2))}+2\pi m_k=}
\label{Betheexpanded}\\
&&\frac{2}{J}\sum_{j\neq k}^S\frac{1}{(x_k-x_j)(1-g^2/(2x_j^2))}
-\frac{2}{J}\frac{g^2}{2x_k^2}\frac{1}{1-g^2/(2x_k^2)}\sum_{j\neq k}^S
\frac{1}{1-g^2/(2x_j^2)} \frac{1}{x_j} \nonumber \\
&& +\frac{1}{J}\mbox{Anomaly}(x_k)+\frac{1}{J}\mbox{Non-analytic}(x_k),
\nonumber
\eea
where $m_k$ is a mode number coming from the ambiguity of
the logarithm. The two first lines constitute the classical
Bethe equations and the last line contains the one-loop correction.
The one-loop correction consists of two terms.
The term $\mbox{Anomaly}(x_k)$ is a spin chain finite size effect. It
arises due to the
fact that the naive expansion of the logarithm becomes invalid
when $x_j-x_k\sim {\cal O}(1/J)$~\cite{Beisert:2005mq}.
This term is {\it analytic} in
$\lambda$. As indicated by the notation, the other one-loop term
is {\it non-analytic} in $\lambda$.
It is
the part of $\theta(x_j,x_k)$
which originates from the $\frac{1}{\sqrt{\lambda}}$
term in eqn.~\rf{crs}, i.e.\ the
Hernandez-Lopez phase~\cite{Hernandez:2006tk}.
Notice that the leading part of $\theta(x_j,x_k)$,
i.e.\ the AFS phase~\cite{Arutyunov:2004vx}, contributes already at the
classical level. Now we make the assumption about the distribution of
Bethe roots that is known to lead to the folded string
solution~\cite{Beisert:2003ea},
namely we assume that the roots lie in two intervals $[-b,-a]$ and
$[a,b]$ on the real axis and are symmetrically distributed around zero.
This means that the
second term on the right hand side of eqn.~\rf{Betheexpanded} 
vanishes.\footnote{The fact that the sum in this term does not include 
the root at
$j=k$ is an $1/J$ effect which can be ignored as the term does not
have any accompanying factors of $\log(\frac{S}{J})$.}
Furthermore, we assign the mode number
$-n$ to roots lying in the right interval and mode number $+n$ to roots
lying in the left interval. Finally, we introduce a resolvent
corresponding to the roots lying in the right interval
\beq
G(x)=\frac{1}{J}\sum_{j=1}^{S/2} \frac{1}{x-x_j}\frac{1}{1-g^2/(2x_j^2)}
\equiv \int_a^b \dd y\, \frac{\rho(y)}{x-y},
\eeq
and we assume that $G(x)$ has a well-defined expansion
in $\frac{1}{J}\sim \frac{1}{\sqrt{\lambda}\log (S/J)}$, i.e.
\beq
G(x)=G_0(x)+\frac{1}{J}G_1(x)+\ldots,
\eeq
where each $G_i(x)$ is analytic in the complex plane except for a
cut $[a,b]$. Accordingly, the density $\rho(x)$ needs to have a
well-defined $\frac{1}{J}$ expansion
\beq
\rho(x)=\rho_0(x)+\frac{1}{J} \rho_1(x)+\ldots
\eeq
with each term in the expansion having support on the interval $[a,b]$.
The normalization condition for $\rho(x)$ reads
\beq
\int_a^b \dd y\, \rho(y)\left\{1-\frac{g^2}{2y^2}\right\}=\frac{S}{2J} \equiv
\frac{\alpha}{2},
\label{fullnormalization}
\eeq
and the string energy, $E$, is encoded in $\rho(y)$ in the following way
\beq
\int_{a}^b \dd y\, \rho(y)=\frac{S}{2J}+\frac{E-S-J}{4J}.
\eeq
If we write
\beq
G(x)=G_+(x)+xG_-(x), \hspace{0.7cm} \mbox{where} \hspace{0.7cm}
G_{\pm}(x)=G_{\pm}(-x),
\eeq
we have
\beq
E=J+S+2Jg^2 \int \dd x \frac{\rho(x)}{x^2}=J+S-2Jg^2 G_-(0).
\label{E}
\eeq
Using the resolvent we can write the Bethe equation in the classical
limit as
\beq
G_0(x+\ii0)+G_0(x-\ii0)-2G_0(-x)=-\frac{1/x}{1-g^2/(2x^2)}+2\pi n,
\hspace{0.7cm}x\in [a,b].
\label{saddle}
\eeq
This equation~\rf{saddle} is nothing but the saddle point equation of the
$O(n)$ model
on a random lattice for $n=-2$~\cite{Kostov:1988fy} with the terms
on the right hand side playing the role of the derivative of the
potential. Its solution with the given
boundary conditions can be written in various
ways~\cite{Kostov:1992pn,Eynard:1995nv}.
Here we shall
use the formulation of~\cite{Eynard:1995nv} where
the solution is given in closed
form for any potential using contour integrals.
In order to find the one-loop correction to the string energy we
have to take into account also the two last terms in eqn.~\rf{Betheexpanded}.
These terms can, at the order considered,
be expressed in terms of the leading order density as follows
\beq
\mbox{Anomaly}(x)= -\frac{1}{1-g^2/(2x^2)}\,(\pi \rho_0'(x))
\left(\coth(\pi \rho_0(x))-\frac{1}{\pi \rho_0(x)}\right),
\eeq
and
\beq
\hbox{Non-analytic}(x)=\frac{1}{\pi} \frac{x^2}{x^2-g^2/2} \,
\int_a^b
\dd y\, \rho_0(y)\left[\Delta \phi(x,y)+\Delta\phi(x,-y)\right]
\label{phase}
\eeq
where
\beq
\Delta \phi(x,y)=
\frac12\sum_{r=2}^{\infty} \sum_{m=0}^{\infty}
c_{r,2m+r+1}^{(1)}\left(\frac{g}{\sqrt{2}}\right)^{2m+2r-1}
\left(\frac{1}{x^{r}y^{2m+r+1}}-\frac{1}{x^{2m+r+1}y^{r}}\right).
\eeq
Notice that we have taken into account the fact that the full
set of Bethe roots is
distributed symmetrically around zero by forming the combination
$\left[\Delta \phi(x,y)+\Delta\phi(x,-y)\right]$.

\section{The semi-classical string energy \label{semiclassical}}

As mentioned above the leading order equation~\rf{saddle} is nothing
but the saddle point equation of the $O(n)$ model on a random
lattice for $n=-2$ and its solution can conveniently
be written down using contour integrals~\cite{Eynard:1995nv}
\bea
G_{0-}(x)&=&\frac{1}{2}
\oint_{{\cal C}_+} \frac{\dd y}{2\pi \ii} \frac{V'_0(y)}{x^2-y^2}
\left\{
\frac{(x^2-a^2)^{1/2}(x^2-b^2)^{1/2}}{(y^2-a^2)^{1/2}(y^2-b^2)^{1/2}}
\right\}, \label{G-}\\
G_{0+}(x)&=& 2\oint_{{\cal C}_+} \frac{\dd y}{2 \pi \ii}G_-(y)
\frac{y^2}{x^2-y^2},
\eea
where the contour encircles ${\cal C}_+=[a,b]$ counterclockwise
and where
\beq
V'_0(y)= -\frac{1/y}{1-g^2/(2y^2)}+2 \pi n.
\label{potential}
\eeq
The endpoints of ${\cal C}_+$,
$a$ and $b$, are determined by
\beq
\oint_{\cal C_+} \frac{\dd y}{2\pi \ii} \frac{V'_0(y)}
{(y^2-a^2)^{1/2}(y^2-b^2)^{1/2}}=0,
\eeq
and
\beq
\oint_{{\cal C}_+} \frac{\dd y}{2\pi \ii} \frac{V'_0(y)y^2}
{(y^2-a^2)^{1/2}(y^2-b^2)^{1/2}}
+\frac{g^2}{2}
\oint_{{\cal C}_+}
\frac{\dd y}{2\pi \ii} \frac{V'_0(y) a b}
{y^2(y^2-a^2)^{1/2}(y^2-b^2)^{1/2}}=\frac{S}{J}.
\eeq
The first condition expresses the fact that $G(x)$ should tend to $0$ as $x$
tends to infinity, and the second condition is a rewriting of eqn. \rf{fullnormalization}.
We need that the Bethe roots stay away from the singularities of the potential,
i.e.\ the points $y=0$ and $y=\pm \frac{g}{\sqrt{2}}$. This means that we
must have $g^2<2a^2$ or $2b^2<g^2$. We choose to work with the former 
assumption, i.e.\
\beq
g^2<2a^2, \label{g2lessthan2a2}
\eeq
 as this will directly reproduce the result of~\cite{Beisert:2003ea}
in the case $g=0$.
Inserting the explicit expression~\rf{potential} for the potential $V_0'(y)$
the boundary conditions read
\bea
0&=&\frac{2n}{b}K(k')-\frac{1}{2}\frac{1}{\sqrt{(a^2-g^2/2)(b^2-g^2/2)}},
\label{boundary1}
\eea
and
\bea
\frac{S}{J}&=&2 nbE(k')-\frac{1}{2}-\frac{1}{2}
\frac{g^2/2}{\sqrt{(a^2-g^2/2)(b^2-g^2/2)}}+
n g^2\frac{1}{a} E(k')
\label{boundary2}
\\
&&
+\frac{1}{2}
\left\{1-\frac{ab}{\sqrt{(a^2-g^2/2)(b^2-g^2/2)}}\right\}, \nonumber
\eea
where $K(k')$ and $E(k')$ are standard elliptic integrals of the first and
the second kind respectively, with $k'$ being given by
\beq
k=\frac{a}{b}, \hspace{0.5cm}
k'=(1-k^2)^{1/2}. \hspace{0.5cm}
\eeq
Furthermore, the expression for the semi-classical
string energy takes the form
\bea
E_0-S-J&=&-a\, b\,J\, g^2 \oint_{{\cal C}_+}\frac{\dd \om}{2 \pi \ii}
\frac{V'(\om)}{\om^2(\om^2-a^2)^{1/2}(\om^2-b^2)^{1/2}}\nonumber \\
&=&-J\,g^2\left\{\frac{2n}{a}\, E(k')+\frac{1}{g^2}
\left[1-\frac{a\,b}{\sqrt{(a^2-g^2/2)(b^2-g^2/2)}}\right]\right\}.
\label{energy}
\eea
Considering only the terms of leading order in $g$ we reproduce the
results of~\cite{Beisert:2003ea},  namely
\beq
a=\frac{1}{4n K(k')},\hspace{0.7cm}
\alpha+\frac{1}{2}=\frac{E(k')}{2K(k')}\frac{1}{k},
\eeq
and
\beq
E_0-S-J=\frac{\lambda\, n^2}{2 \pi^2 J} K(k') \left\{(1+k^2)K(k')-2E(k')\right\}.
\eeq
It is obvious that by means of the equations~\rf{boundary1},
~\rf{boundary2} and~\rf{energy} one can recursively express the
semi-classical
energy order by order in $\lambda$. This idea has been pursued f.inst.
in references~\cite{Belitsky:2006en,Basso:2006nk}.
Here, we shall in stead consider the limit~\rf{limit}
\beq
1\ll J\ll S, \hspace{0.5cm} z=\frac{\sqrt{\lambda }\;n }{\pi J}
\log\left(\frac{S}{J}\right)\hspace{0.5cm}{\mbox{fixed}},
\label{limit3}
\eeq
where it is possible
to obtain a closed expression for the all-loop energy.
We immediately see that in this limit we have
\beq
k\rightarrow 0, \hspace{0.5cm} a\rightarrow 0,\hspace{0.5cm}
b\rightarrow \infty,
\eeq
and from the second boundary equation~\rf{boundary2} we find
\beq
\log\left(\frac{S}{J}\right)\sim K(k')\sim \log\left(\frac{1}{k}\right).
\eeq
Introducing the notation
\beq
{\hat{g}}=\frac{g}{\sqrt{2} a}, \label{ghat}
\eeq
we notice that the first boundary boundary equation~\rf{boundary1} can be
written as
\beq
a=\frac{1}{4n \sqrt{1-\hat{g}^2}\sqrt{1-k^2\hat{g}^2}K(k')},
\label{ahat}
\eeq
and therefore in the limit~\rf{limit3} we have
\beq
\hat{g}^2=\frac{z^2}{z^2+1}, \label{gz}
\eeq
and in particular $g^2 <2a^2$.
Using eqn.~\rf{ahat} we can express the energy as
\beq
E_0=S+J\frac{1}{\sqrt{(1-\hat{g}^2)(1-k^2\hat{g}^2)}}
\left[1-\hat{g}^2\frac{E(k')}{K(k')}\right].
\eeq
{}From here we immediately find, in the limit given by eqn.~\rf{limit3},
\beq
E_0=S+J\sqrt{1+z^2}, \label{squareroot}
\eeq
which agrees exactly with the string theory result of
reference~\cite{Frolov:2006qe}, cf.\ eqn~\rf{E0}.
{}From our result for $G_{0-}$ we can extract the Bethe root distribution
at leading order $\rho_0(x)$ in terms of which the one loop correction terms
are expressed. One finds
\bea
\rho_0(x)&=&\frac{x}{i\pi}(G_{0-}(x-\ii0)-G_{0-}(x+\ii0)) \nonumber \\
&=&
-\frac{x}{\pi}(x^2-a^2)^{1/2}(b^2-x^2)^{1/2}
\oint_{{\cal C}_+} \frac{\dd y}{2\pi \ii} \frac{V'(y)}{x^2-y^2}
\left\{
\frac{1}{(y^2-a^2)^{1/2}(y^2-b^2)^{1/2}}
\right\} \nonumber \\
&=&-\frac{x}{\pi}(x^2-a^2)^{1/2}(b^2-x^2)^{1/2}
\int_{a}^b \hspace{-0.55cm} - \hspace{0.35cm}
\frac{\dd y}{\pi} \frac{V'(y)}{x^2-y^2}
\left\{
\frac{1}{(y^2-a^2)^{1/2}(b^2-y^2)^{1/2}}
\right\} \nonumber \\
&=&\frac{x\sqrt{x^2-a^2}}{2\pi\,b\,\sqrt{b^2-x^2}}
\left(\frac{b}{x^2-\frac{g^2}{2}}\frac{\sqrt{b^2-\frac{g^2}{2}}
   }{\sqrt{a^2-\frac{g^2}{2}}}
   -4\, n\, \Pi
   \left(1-\frac{x^2}{b^2},k'\right)\right),\label{fulldensity}
\eea
where in the last integral the principal value appears.
Considering only leading order in $g$ and setting $n=1$ we recover
the  expression obtained in~\cite{Beisert:2003ea}.
Introducing
\beq
\om=\frac{x}{a},
\eeq
we can also write
\bea
\rho_0(\omega)&=&
 \frac{2 \,k\, n\, \omega\,  \sqrt{\omega ^2-1}}{\pi\,
   \sqrt{\frac{1}{k^2}-\omega ^2}}\left(\frac{\frac{1}{k^2}-\hat{g}^2}
   {\omega ^2-\hat{g}^2}\,K(k')-\Pi \left(1-k^2 \omega ^2,k'\right)\right),
\eea
where now the normalization condition reads
\beq
\int_a^b\dd x\, \rho_0(x)\left\{1-\frac{g^2}{2\,y^2}\right\}
= a\int_1^{1/k} \dd\om\, \rho_0(\om)\left\{1-\frac{\hat{g}^2}{\om^2}\right\}=\frac{S}{2J}.
\eeq
We also note the formula
\bea
\lefteqn{
\rho_0'(\omega)=
\frac{2\,n }{k\, \pi
\,\sqrt{\frac{1}{k^2}-\omega ^2}\,\sqrt{\omega ^2-1}}\;\times}\\
&&
\left\{\left(k^2 \hat{g}^2+\frac{\left(1+\left(1-2 \hat{g}^2\right) k^2\right)
   \hat{g}^2}{\hat{g}^2-\omega ^2}+\frac{\left(1-\hat{g}^2\right) \left(1-k^2
   \hat{g}^2\right) \left(\hat{g}^2+\omega ^2\right)}{\left(\hat{g}^2-\omega
   ^2\right)^2}\right) K(k')-E(k')\right\}.\nonumber \\
   && \nonumber
\eea
Taking the limit~\rf{limit}, we get
\bea
\rho_0(\om)&\approx&\frac{n}{\pi} \frac{\sqrt{\om^2-1}}{\om}
\left\{2\log(\alpha) \sqrt{1-k^2\om^2}\,
\frac{\hat{g}^2}{\om^2-\hat{g}^2}+
\log\left[\frac{1+\sqrt{1-k^2\om^2}}{1-\sqrt{1-k^2 \om^2}} \right]
\right\}
\nonumber \\
&\approx
& \frac{2n}{\pi} \frac{\om \sqrt{\om^2-1}}{\om^2-\hat{g}^2} \log(\alpha),
\label{rho0}
\eea
where the latter expression of course needs to be treated with some care.
Furthermore,
\beq
\rho_0'(\om)=\frac{2n}{\pi} \frac{(1-2\hat{g}^2)\om^2+\hat{g}^2}
{\sqrt{\om^2-1}(\om^2-\hat{g}^2)^2}
\log(\alpha).
\label{rho0prime}
\eeq
\section{The one-loop string energy\label{oneloop}}
Including the one-loop corrections, our Bethe equations read
\beq
G(x+\ii0)+G(x-\ii0)-2G(-x)=V_0'(x)+\frac{1}{J}V_1'(x),
\eeq
with $V_0'(x)$ given by eqn.~\rf{potential} and with
\beq
V_1'(x)=\mbox{Anomaly}(x)+\mbox{Non-analytic}(x).
\eeq
By applying the solution formula~\rf{G-} to this equation and expanding
everything including the interval boundaries in $\frac{1}{J}$, one derives
the following formula for $G_{1-}(x)$
\beq
G_{1-}(x)=\frac{1}{2}
\oint_{{\cal C}_+} \frac{\dd\omega}{2\pi \ii}V_1'(\om)
\left(\frac{1}{x^2-\om^2} -
\frac{\frac{g^2}{2ab}}{1-\frac{g^2}{2 a b}}\frac{1}{\om^2}
\right)
\left\{
\frac{(\om^2-a^2)^{1/2}(\om^2-b^2)^{1/2}}{(x^2-a^2)^{1/2}(x^2-b^2)^{1/2}}
\right\}, \label{G1-}
\eeq
where we stress that the points $a$ and $b$ are the same as for the leading
order solution. The one-loop contribution to the energy then reads,
cf.\ eqn.~\rf{E}
\beq
E_1=-2 g^2 G_{1-}(0). \label{E1simple}
\eeq

\subsection{The spin chain finite size correction\label{finitesize}}

As explained above the spin chain finite size corrections will give us
the analytic part of the one-loop string energy. This contribution
is determined from~\rf{E1simple}
by inserting $\mbox{Anomaly}(x)$ at the place of $V_1'(x)$ in
eqn.~\rf{G1-}. One gets
\bea
\lefteqn{(E_1)^{{\hbox{\scriptsize analytic}}}_{{\hbox{\scriptsize Bethe}}}= }
\nonumber\\
&&\mbox{}-\frac{\frac{g^2}{ab}}{1-\frac{g^2}{ab}}
\int_a^b \frac{\dd x}{\pi} \frac{1}{1-\frac{g^2}{2x^2}}
(\pi \rho_0'(x)) 
\left(\coth(\pi \rho_0(x))-\frac{1}{\pi\rho_0(x)}\right) \frac{\sqrt{(x^2-a^2)(b^2-x^2)}}{x^2}.
\nonumber
\eea
In the limit we are interested in, $\rho_0(x)$ and $\rho_0'(x)$ are given
by eqns.~\rf{rho0} and~\rf{rho0prime}. In particular, since $\rho_0(x)$
contains the divergent factor $\log( \alpha)$ we can use the approximation
$\coth(\pi \rho_0(x))-\frac{1}{\pi \rho_0(x)}=1$. In this way the integral above becomes
\bea
(E_1)^{{\hbox{\scriptsize analytic}}}_{{\hbox{\scriptsize Bethe}}}&=&
-\frac{g^2}{a^2} (2n \log(\alpha)) \int_1^{\infty}
\frac{\dd\omega}{\pi}\frac{(1-2\hat{g}^2)\om^2+\hat{g}^2}{(\om^2-\hat{g}^2)^3}
 \\
&=& -4\frac{J}{\sqrt{\lambda}} z^3 (1-\hat{g}^2)\int_1^{\infty}
\frac{\dd\omega}{\pi}\frac{(1-2\hat{g}^2)\om^2+\hat{g}^2}{(\om^2-\hat{g}^2)^3} \\
&=& \frac{J}{\sqrt{\lambda}} \left(
z-\frac{1+2z^2}{\sqrt{1+z^2}}\log \left[ z+\sqrt{1+z^2}\right]\right)
\label{analyticBethe}
\eea
which exactly agrees with the expression~\rf{analytic1} obtained in
reference~\cite{Frolov:2006qe}.

\subsection{The HL phase \label{HLphase}}

The non-analytic contribution is given by the 
the HL phase (\ref{AFSphase}) through the coefficients $c_{r,s}^{(1)}$
of (\ref{coefc1}). More precisely,
\beq
\hbox{Non-analytic}(x)=\frac{1}{\pi} \frac{x^2}{x^2-g^2/2} \,
\int_{a}^b
\dd y\, \rho_0(y)\left[\Delta \phi(x,y)+\Delta \phi(x,-y)\right],
\label{phase}
\eeq
where
\beq
\Delta \phi(x,y)=
\frac12\sum_{r=2}^{\infty} \sum_{m=0}^{\infty}
c_{r,2m+r+1}^{(1)}\left(\frac{g}{\sqrt{2}}\right)^{2m+2r-1}
\left(\frac{1}{x^{r}y^{2m+r+1}}-\frac{1}{x^{2m+r+1}y^{r}}\right).
\eeq
Notice that we have taken into account the fact that the full set of 
Bethe roots is symmetrically distributed around zero by forming the
combination $\left[\Delta \phi(x,y)+\Delta \phi(x,-y)\right]$.
Let us define
\beq
\omega=\frac xa\ ,\quad\quad\quad\quad \nu=\frac ya\,,
\eeq
The double sum above can be carried out explicitly and gives
$$\Delta \phi(\omega,\nu)=-\frac1{a^2}
\left\{  \frac{2\,\hat{g}}{(\nu -\omega ) \left(\nu \, \omega -\hat{g}^2\right)}
+
\left(\frac{1}{(\nu -\omega )^2}+\frac{\hat{g}^2}{\left(\nu \, \omega -\hat{g}^2\right)^2}\right)
\log\left[\frac{(\omega -\hat{g})(\nu+\hat{g} )}{(\omega +\hat{g})(\nu -\hat{g})}\
\right]\right\}.$$
Furthermore,
\bea
\lefteqn{
\Delta \phi(\omega,\nu)+\Delta \phi(\omega,-\nu)= } \\
&&-\frac4{a^2}
\left\{\frac{\hat{g} \left(\hat{g}^2+\nu ^2\right) \omega }{\left(\nu ^2-\omega ^2\right) \left(\nu
   ^2 \omega ^2-\hat{g}^4\right)}
+\omega\,\nu\,
\left(\frac{1}{\left(\omega ^2-\nu
   ^2\right)^2}+\frac{\hat{g}^4}{\left(\nu ^2 \omega ^2-\hat{g}^4\right)^2}\right)
\log\left[\frac{\nu+\hat{g}}{\nu -\hat{g}}\right] \right.\nnb\\[3mm]
&&\hspace{1cm}+ \left.
\frac{1}{2} \left(\frac{\omega ^2+\nu ^2}{\left(\omega ^2-\nu ^2\right)^2}+
\frac{\hat{g}^2\left(\nu ^2 \omega ^2+\hat{g}^4\right) }{\left(\nu ^2 \omega
   ^2-\hat{g}^4\right)^2}\right) \log
   \left[\frac{\omega -\hat{g}}{\omega+\hat{g} }\right]\right\}.
\eea
The correction to the energy (\ref{E1simple}) is then given by
\bea
\left(E_1\right)^{\hbox{\scriptsize non-analytic}}_{\hbox{\scriptsize Bethe}}&=&
-2\,a\,\frac{k\,\hat{g}^2}{1-k\,\hat{g}^2}\,\oint_{{\cal C}^+}\frac{\dd\omega}{2\,\ii\,\pi}\,
\frac{\sqrt{\omega ^2-1}
 \sqrt{\omega ^2-\frac{1}{k^2}}}{\omega ^2}\times
\hbox{Non-analytic}(\omega).\qquad.\quad\phantom{1}\label{eqE1na}
\eea
In the limit (\ref{limit}) and in the variables used here, the contour ${\cal C}^+$
transforms into the real half line
$[1,+\infty[$. The non-analytic part of the energy will therefore
be given by the following double integral :
\bea
\lefteqn{
\left(E_1\right)^{\hbox{\scriptsize non-analytic}}_{\hbox{\scriptsize Bethe}}=}
\label{dbint}\\
&&\frac{4\,n\,a^2}{\pi^2}\,\log \left(\alpha\right)\,\hat{g}^2\,
\oint_{{\cal C}^+}\frac{\dd\omega}{2\,\ii\,\pi}\,
\frac{\sqrt{1-\omega ^2}}{\omega^2-\hat{g}^2} \,
\int_1^\infty\dd \nu\, \frac{ \nu\,  \sqrt{\nu ^2-1}}{\nu ^2-\hat{g}^2}
\left[\Delta
 \phi(\omega,\nu)+\Delta\phi(\omega,-\nu)\right].\nonumber 
\eea
This integration is carried out in the Appendix. The result reads
\bea
\lefteqn{
\left(E_1\right)^{\hbox{\scriptsize non-analytic}}_{\hbox{\scriptsize Bethe}}
=  \label{E1na} 
-\frac{n \log (\alpha )}{2 \pi  \hat{g}}\left(2 \hat{g}^2+\left(3-\hat{g}^2\right) \log
   \left(1-\hat{g}^2\right)+\left(1+\hat{g}^2\right) \log \left(1+\hat{g}^2\right)\right) }
\\
&&\hspace{0.3cm}=\frac{J}{\sqrt{\lambda }}\frac{1}{\sqrt{1+z^2}}
\left(-z^2+2 \left(1+z^2\right) \log \left(1+z^2\right)-\left(1+2 z^2\right) \log \left(\sqrt{1+2
   z^2}\right)\right)\nnb
\eea
Here again, our result matches perfectly with the 
expression~\rf{nonanalytic1} from
reference~\cite{Frolov:2006qe}.

\section{Conclusion \label{Conclusion}}

We have extracted the strong coupling limit of the scaling function for low 
twist, high spin operators of ${\cal N}=4$ SYM from the quantum string Bethe
equations by applying these to a folded string rotating  with angular 
momenta $(S,J)$ in 
$AdS_3\times S^1\subset AdS_5\times S^5$ and considering the limit
\beq
1\ll J \ll S, \hspace{0.5cm} z=\frac{\sqrt{\lambda}}{\pi J} 
\log\left(\frac{S}{J}\right), \hspace{0.5cm} \mbox{fixed}. 
\eeq
It is interesting to notice that this limit which 
was observed 
in~\cite{Belitsky:2006en} and further explored in~\cite{Frolov:2006qe} 
from the string theory perspective  
also follows naturally from the quantum string Bethe ansatz. Namely, 
assuming the simplest possible analyticity structure with two cuts one
is led to the relation~\rf{g2lessthan2a2} and using the rewritings in
eqns.~\rf{ghat} to~\rf{gz} the quantity $z$ naturally appears.

Our computation involved first a solution of the Bethe equations at
the classical level. This part was straightforward and immediately led
to the square root expression~\rf{squareroot} for the classical energy.
Subsequently, we determined the one-loop contribution to the energy.
This contribution consisted of two parts, one originating from spin 
chain finite size effects and one being due to the Hernandez-Lopez
phase. Both parts could be treated exactly and led to a total expression
for the string one-loop energy, $\frac{J}{\sqrt{\lambda}}F(z)$, 
which agreed at the functional level
with the result of a traditional string theory computation, cf.\ eqns. 
\rf{analytic1}, \rf{nonanalytic1}, \rf{analyticBethe} and~\rf{E1na}. 
Both the classical and the one loop energy when
considered as a function of $z$ could be smoothly extrapolated to 
large values of $z$ and led to the strong coupling limit of the scaling
function
\beq
f(\lambda)=\frac{\sqrt{\lambda}}{\pi}-\frac{3\log(2)}{\pi} +
{\cal O}\left(\frac{1}{\sqrt{\lambda}}\right) . \label{scalingconcl}
\eeq
We stress again that the famous $\frac{-3\log(2)}{\pi}$ is due to
a highly non-trivial cancellation between terms originating from the
HL-phase and terms due to spin chain finite size effects. 
More precisely, we have 
\bea
E_1^{\hbox{\scriptsize analytic}}& \sim & 
\left(\frac{-2\log(z)+1-2\log(2)}{\pi}\right)
\log\left(\frac{S}{J}\right)
\hspace{0.7cm} \mbox{as} \hspace{0.7cm}
z\rightarrow \infty,\\
E_1^{\hbox{\scriptsize non-analytic}} &\sim& \left(\frac{2\log(z)-1-\log(2)}{\pi}\right)
\log\left(\frac{S}{J}\right)
\hspace{0.7cm} \mbox{as} \hspace{0.7cm}
z\rightarrow \infty.
\eea
As mentioned
earlier there exists a numerical prediction for the coefficient of 
the ${\cal O}(1/{\sqrt{\lambda}})$ term 
of~\rf{scalingconcl}~\cite{Benna:2006nd}. 
Furthermore, a genuine string theory calculation of the same quantity
seems to be under way~\cite{Roiban:2007jf}. Given these developments 
it might be interesting to pursue our approach to two-loop order. It
is obvious that the same strategy should be applicable and we are convinced
that the Bethe equations will once again prove their efficiency.

\vspace*{1.0cm}

\noindent
{\bf Acknowledgments}
The authors thank Lisa Freyhult, Sergey Frolov and Matthias Staudacher
for useful discussions.
Both authors were supported by
ENRAGE (European Network on Random Geometry), a Marie Curie
Research Training Network financed by the European Community's
Sixth Framework Programme, network contract MRTN-CT-2004-005616.

\begin{appendix}
\section*{Appendix}
\begin{figure}[htbp]
\begin{center}
   \psfrag{a}{$0$}
   \psfrag{b}{$\hat{g}$}
   \psfrag{c}{-$\hat{g}$}
   \psfrag{d}{$1$}
   \psfrag{e}{$-1$}
   \psfrag{f}{${\cal C}_0$}
   \psfrag{g}{${\cal C}_+$}
   \psfrag{h}{${\cal C}_-$}
   \includegraphics[height=2.2cm]{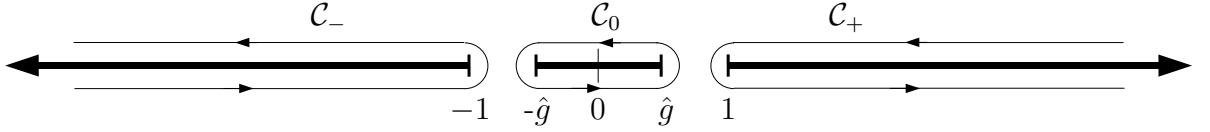}
\end{center}
  \caption{Cuts on the complex plane in the integration of ${\cal B}(\omega,\nu)$.}\label{fig2}
\end{figure}

In this appendix, we present the evaluation of the 
double integral (\ref{dbint})
which up to a factor $-\frac{16\,n\,\hat{g}^2}{\pi^2}\,\log(\alpha)$ can be written as
\beq
{\cal I}=
\oint_{{\cal C}^+}\frac{\dd\omega}{2\,\ii\,\pi}\,
\frac{\sqrt{1-\omega ^2}}{\omega^2-\hat{g}^2} \,
\int_1^\infty\dd \nu\, \frac{ \nu\,  \sqrt{\nu ^2-1}}{\nu ^2-\hat{g}^2}
\left({\cal A}(\omega,\nu)+{\cal B}(\omega,\nu)\right),\label{integral}
\eeq
where
\bea
{\cal A}(\omega,\nu)&=&\frac{\hat{g} \left(\hat{g}^2+\nu ^2\right) \omega }
{\left(\nu ^2-\omega ^2\right) \left(\nu
   ^2 \omega ^2-\hat{g}^4\right)}
+\omega\,\nu\,
\left(\frac{1}{\left(\omega ^2-\nu
   ^2\right)^2}+\frac{\hat{g}^4}{\left(\nu ^2 \omega ^2-\hat{g}^4\right)^2}\right)
\log\left[\frac{\nu+\hat{g}}{\nu -\hat{g}}\right]~,\nnb\\[3mm]
{\cal B}(\omega,\nu)&=&
\frac{1}{2} \left(\frac{\omega ^2+\nu ^2}{\left(\omega ^2-\nu ^2\right)^2}+
\frac{\hat{g}^2\left(\nu ^2 \omega ^2+\hat{g}^4\right) }{\left(\nu ^2 \omega
   ^2-\hat{g}^4\right)^2}\right)\,\log
   \left[\frac{\omega -\hat{g}}{\omega+\hat{g} }\right] ~,\nnb
\eea
and the contour ${\cal C}^+$ is the real half line
$[1,+\infty[$.
The integration of ${\cal A}(\omega,\nu)$ with
respect to $\omega$ is straightforward :
\bea
\oint_{{\cal
C}^+}\frac{\dd\omega}{2\,\ii\,\pi}\frac{\sqrt{1-\omega ^2}
}{\omega^2-\hat{g}^2} {\cal A}(\omega,\nu)&=&-\mathop{\rm Res}_{\tiny\omega=\hat{g}}
       \left[\frac{\sqrt{1-\omega ^2} }{\omega^2-\hat{g}^2}  {\cal A}(\omega,\nu)\right]
       -\mathop{\rm Res}_{\tiny\omega=\ft{\hat{g}^2}{\nu}}
       \left[\frac{\sqrt{1-\omega ^2} }{\omega^2-\hat{g}^2}  {\cal A}(\omega,\nu)\right]\nnb\\
&=&-\frac{\left(\hat{g}^2+\nu ^2\right)\sqrt{1-\hat{g}^2} -\nu  \sqrt{\nu
   ^2-\hat{g}^4}}{2\,  \hat{g}\, \left(\hat{g}^2-\nu
   ^2\right)^2}\label{AA}\\
&&  -\frac{\left(\hat{g}^4+\nu ^2 \hat{g}^2-2 \nu ^2+4 \sqrt{1-\hat{g}^2} \nu
   \sqrt{\nu ^2-\hat{g}^4}\right) }{4 \,\left(\hat{g}^2-\nu ^2\right)^2 \sqrt{\nu ^2-\hat{g}^4}}
   \log \left(\frac{\nu+\hat{g} }{\nu -\hat{g}}\right)\nnb
\eea
In order to integrate the last term of (\ref{integral}), one first 
exploits the parity of the integrand to 
extend the contour ${\cal C}_+$ to ${\cal C}_+\cup{\cal C}_-$ 
(see Figure \ref{fig2}). Subsequently, the resulting contour 
can be deformed into the contour ${\cal C}_0$ around the cut $[-\hat{g}, \hat{g}]$. This
contour integral can then be re-expressed as the finite part of the
integral along $[-\hat{g}, \hat{g}]$  with the substitution
$$\log\left[\frac{\omega -\hat{g}}{\omega+\hat{g} }\right]\longrightarrow2\,\ii\,\pi~.$$

One gets
\bea
\oint_{{\cal C}^+}\frac{\dd\omega}{2\,\ii\,\pi}\frac{\sqrt{1-\omega ^2} }{\omega^2-\hat{g}^2}
\,{\cal B}(\omega,\nu)
&=&
-\frac12 \oint_{{\cal C}_0}
                \frac{\dd\omega}{2\,\ii\,\pi}\frac{\sqrt{1-\omega ^2} }{\omega^2-\hat{g}^2}
                \,{\cal B}(\omega,\nu)\\[3mm]
&=&\frac12
         \int_{-\hat{g}}^{\hat{g}}\hspace{-0.7cm} -\ \ \dd\omega
\frac{\sqrt{1-\omega ^2} }{\omega^2-\hat{g}^2}
\,\frac{1}{2} \left(\frac{\omega ^2+\nu ^2}{\left(\omega ^2-\nu ^2\right)^2}+
\frac{\hat{g}^2\left(\nu ^2 \omega ^2+\hat{g}^4\right) }{\left(\nu ^2 \omega
   ^2-\hat{g}^4\right)^2}\right)\\[3mm]
&=&\frac{\left(\hat{g}^2+\nu ^2\right)\sqrt{1-\hat{g}^2} -\nu  \sqrt{\nu
   ^2-\hat{g}^4}}{2\,  \hat{g}\, \left(\hat{g}^2-\nu
   ^2\right)^2}\nnb\\
&& -\frac{\nu  \left(\hat{g}^2+\nu ^2-2\right)}{2\,\left(\nu^2-\hat{g}^2\right)^2\,
   \sqrt{\nu ^2-1}} \tan ^{-1}\left(\frac{\hat{g} \sqrt{\nu
   ^2-1}}{\nu\sqrt{1-\hat{g}^2} }\right)\nnb\\
&&+\frac{\hat{g}^4+\left(\hat{g}^2-2\right) \nu ^2}{2\,\left(\nu^2-\hat{g}^2\right)^2\,\sqrt{\nu ^2-\hat{g}^4}}
  \tanh ^{-1}\left(\frac{\hat{g}^3}{\nu +\sqrt{1-\hat{g}^2} \sqrt{\nu ^2-\hat{g}^4}}\right)\nnb\\
&&-\frac{\sqrt{1-\hat{g}^2} \left(\hat{g}^2+\nu ^2\right)}{2\,\left(\nu^2-\hat{g}^2\right)^2\,\hat{g}}
  \log \left(1-\hat{g}^2\right)~.\label{BB}
\eea
Here, $\displaystyle\int_{-\hat{g}}^{\hat{g}}\hspace{-0.7cm} -\ \ $ means that all poles in the
interval $[-\hat{g},\hat{g}]$ should
be subtracted from the integrand.
Summing (\ref{AA}) and (\ref{BB}), we obtain
\bea
\oint_{{\cal C}^+}\frac{\dd\omega}{2\,\ii\,\pi}\frac{\sqrt{1-\omega ^2} }{\omega^2-\hat{g}^2}
\,\left({\cal A}(\omega,\nu)+{\cal B}(\omega,\nu)\right)
&=&-\frac{\sqrt{1-\hat{g}^2} \left(\hat{g}^2+\nu ^2\right)}
         {2\left(\nu ^2-\hat{g}^2\right)^2\,\hat{g}}
    \log \left(1-\hat{g}^2\right)\nnb\\
&&-\frac{2 \sqrt{1-\hat{g}^2} \nu}{\left(\nu ^2-\hat{g}^2\right)^2\,}
    \tanh ^{-1}\left(\frac{\hat{g}}{\nu }\right)\nnb\\
&& -\frac{\nu  \left(\hat{g}^2+\nu ^2-2\right)}
         {2\left(\nu ^2-\hat{g}^2\right)^2\,\sqrt{\nu ^2-1}}
     \tan ^{-1}\left(\frac{\hat{g} \sqrt{\nu ^2-1}}{\nu \sqrt{1-\hat{g}^2} }\right)\nnb\\
&&-\frac{\hat{g}^4+\left(\hat{g}^2-2\right) \nu ^2}
        {2\left(\nu ^2-\hat{g}^2\right)^2\,\sqrt{\nu ^2-\hat{g}^4}}
   \tanh ^{-1}\left(\frac{\hat{g} \sqrt{1-\hat{g}^2}}{\sqrt{\nu ^2-\hat{g}^4}}\right).\nnb\\\label{intomega}
\eea
The next task is to integrate (\ref{intomega}) with respect to $\nu$.
This can be done using the same kind of techniques as previously after rewriting the integral
as a contour integral around ${\cal C}^+$. By this strategy, one gets the following intermediate results :

\bea
\int_1^{+\infty}\frac{\nu\   \sqrt{\nu ^2-1}}{\nu ^2-\hat{g}^2}
   \frac{\hat{g}^2+\nu ^2}{\left(\nu ^2-\hat{g}^2\right)^2} \, \dd\nu
   =\frac{\pi}{8}\,\frac{2-\hat{g}^2}{\left(1-\hat{g}^2\right)^{3/2}}\ ,\hspace*{6.5cm}
\eea
\bea
\int_1^{+\infty}\frac{\nu\   \sqrt{\nu ^2-1}}{\nu ^2-\hat{g}^2}
   \frac{\nu}{\left(\hat{g}^2-\nu ^2\right)^2}\tanh ^{-1}\left(\frac{\hat{g}}{\nu }\right)\, \dd\nu\hspace*{6.7cm}\nnb\\
=-\frac{\pi}{32 \,\hat{g}\,
   \left(1-\hat{g}^2\right)^{3/2}}\left(1-\frac{3}{2}\, \hat{g}^2
   +\frac{1}{\hat{g}^2}\,\log
   \left(1-\hat{g}^2\right)\right)\ ,
\eea
\bea
\int_1^{+\infty} \frac{\nu\   \sqrt{\nu ^2-1}}{\nu ^2-\hat{g}^2}
   \frac{\nu  \left(\hat{g}^2+\nu ^2-2\right) }{\left(\hat{g}^2-\nu ^2\right)^2
   \sqrt{\nu ^2-1}}\,\tan ^{-1}\left(\frac{\hat{g}}{\nu}
   \frac{\sqrt{\nu ^2-1}}{\sqrt{1-\hat{g}^2}}\right) \dd\nu\hspace*{4cm}\nnb\\
=-\frac{\pi}{16\, \hat{g} \left(1-\hat{g}^2\right)}
   \left(1-\frac{3 }{2}\,\hat{g}^2+\frac{1-\hat{g}^4}{\hat{g}^2}\log
   \left(1-\hat{g}^2\right)\right)\ ,
\eea
\bea
\int_1^{+\infty} \frac{\nu\   \sqrt{\nu ^2-1}}{\nu ^2-\hat{g}^2}
   \frac{\left(\hat{g}^4+\left(\hat{g}^2-2\right) \nu ^2\right) \tanh ^{-1}\left(\frac{\hat{g}
   \sqrt{1-\hat{g}^2}}{\sqrt{\nu ^2-\hat{g}^4}}\right)}{\left(\hat{g}^2-\nu ^2\right)^2
   \sqrt{\nu ^2-\hat{g}^4}} \dd\nu\hspace*{4.1cm}\nnb\\
=\frac{\pi }{16\, \hat{g}\, \left(1-\hat{g}^2\right)}
   \left(1-\frac{5}{2}\, \hat{g}^2-\frac{1-\hat{g}^4 }{\hat{g}^2}\log
   \left(1+\hat{g}^2\right)\right).
\eea
Finally, putting everything together we end up with
\bea
{\cal I}&=&\frac{\pi }{32 \hat{g}^3}\left(2 \hat{g}^2+\left(3-\hat{g}^2\right) \log
   \left(1-\hat{g}^2\right)+\left(1+\hat{g}^2\right) \log \left(1+\hat{g}^2\right)\right)~,
\eea
which, up to the factor $-\frac{16\,n\,\hat{g}^2}{\pi^2}\,\log(\alpha)$, gives back the result (\ref{E1na}).
\end{appendix}

\end{document}